\documentstyle[12pt]{article}
\begin{document}
%\begin{flushright} QUANT-PH.....
%\end{flushright}

\centerline{{\Large\bf "Structure and  function"}}

\footnote{To appear in Proceed. of The Conference 
          "Towards a Science of Consciousness", Tucson (Az), April 1996}

\vspace{.3in}
\centerline{\bf Giuseppe Vitiello}
 
\vspace{.2in}
\centerline{{\it Dipartimento di Fisica}}
\centerline{{\it  Universit\`a di Salerno, 84100 Salerno, Italy}}
\centerline{{\it  vitiello@vaxsa.csied.unisa.it}}

\begin{abstract}
I discuss the role of quantum dynamics in brain and living matter
physics. The paper is presented in the form of a letter to Patricia
S. Churchland.
\end{abstract}

%\newpage

\medskip
\bigskip
\bigskip

Dear Patricia, after your talk in Tucson I said to myself: "I must meet 
Patricia Churchland and discuss with her on the role of Quantum Mechanics 
(QM) and quantum formalisms in Consciousness studies". However, the 
Conference was very dense, you very busy and I was "not so sure..." from  
where to start discussing with you. So, at the end I decided to write you a 
letter.

In your talk, which I enjoyed a lot, you were keeping saying "I am not so 
sure...", "I am not so sure...". You explained very well why one should 
have real doubts about "hard" (and easy!) problems (on which I will not say 
anything in this letter) and especially about using QM in the study of 
Consciousness.

From what you were saying I realized that you were completely right: "if" 
QM is what you were referring to, and "if" its use and purpose are the ones 
you were saying, "then" your doubts are really sound and, even more, I 
confirm to you that QM is completely useless in Consciousness studies; the 
popular expression: "a mystery cannot solve another mystery" would  be the 
fitting one.

However, as a physicist I want to tell you that one should not talk much 
"about" QM. Physicists, and other scientists as chemists, engineers, etc., 
"use" QM in a large number of practical applications in solid state 
physics, electronics, chemistry, etc. with extraordinary success: it is an 
undeniable fact that our every day (real!) life strongly depends on those 
successful applications of QM; everything is around us (including 
ourselves!) is made of atoms and the Periodic Table of the Elements is 
clearly understood in terms of QM (recall, e.g., the Pauli Principle in 
building up electronic shells in the atoms). QM is not a mystery, from this 
perspective. The photoelectric cell of our elevator or our CD or computer 
have nothing counter-intuitive. Of course, I am not saying that the success 
of QM by itself justifies the use of QM in Consciousness studies. I will 
come back to this point later on.

What I want to stress here is that QM is NOT the OBJECT of our discussion! 
There are certainly many open problems in the interpretation of certain 
aspects of QM which are of great epistemological and philosophical 
interest. However, these problems absolutely do not interfere or diminish 
the great successes of QM in practical applications. It is certainly 
interesting to study these interpretative problems, BUT they are NOT the 
object of our present discussion.

And, please notice that here I am not defending QM, since as I have clearly 
stated many times in my papers, QM does not provide the proper mathematical 
formalism for the study of living matter physics.
The proper mathematical formalism in such a study turns out to be indeed 
the one of Quantum Field Theory (QFT). But this is a too strong statement 
at this moment of our discussion. Let me go by small steps, instead.

I must confess to you that I am not prepared to take as the object of our 
discussion how to approach to the study of Consciousness. As a physicist, I 
would better start by considering some more "material" object, as the brain 
itself or more generally living matter, for example the cell. Here I need 
to explain better myself since the word "material" may be misleading.

In Physics it is not enough to search what things are "made of". Listing 
elementary "components" is a crucial step, but it is only one step. We want 
to know not only what things are made of but ALSO "how all of it works": we 
are interested in the Dynamics. In short, fancy words: we are interested 
"in structures AND in functions"; and we physicists are attached to our 
fixations in a so narcissistic way that we even mix up structure and 
function up to the point that we do not anymore make a sharp distinction 
between them. So, to us, having a detailed list of components does not mean 
to know much about the system under study. Moreover, it is not even 
possible to make a "complete" list of components without knowing how they 
work all together in the system. The same concept of component is 
meaningless outside a "dynamical" knowledge of the system. Thus when I say 
"material" I refer also to dynamical laws, not only to the mere collection 
of components.

After all, what I am saying is quite simple: everybody agrees indeed that 
studying the Tucson phone book does not mean to know the city of Tucson. 
Let me give one more specific physical example: the crystal.

As well known, when some kind of atoms (or molecules) sit in some lattice 
sites we have a crystal. The lattice is a specific geometric arrangement 
with a characteristic length (I am thinking of a very simple situation 
which is enough for what I want to say). A crystal may be broken in many 
ways, say by melting it at high temperature. Once the crystal is broken, 
one is left with the constituent atoms. So the atoms may be in the crystal 
phase or, e.g. after melting, in the gaseous phase. We can think of these 
phases as the functions of our structure (the atoms): the crystal function, 
the gaseous function. In the crystal phase one may experimentally study the 
scattering of, say, neutrons on phonons. Phonons are the quanta of the 
elastic waves propagating in the crystal. They are true particles living in 
the crystal. We observe them indeed in the scattering with neutrons. As 
matter of fact, for the complementarity principle, they are the same thing 
as the elastic waves: they propagate over the whole system as the elastic 
waves do (for this reason they are also called collective modes). The 
phonons (or the elastic waves) are in fact the messengers exchanged by the 
atoms and are responsible for holding the atoms in their lattice sites. 
Therefore the list of the crystal components includes not only the atoms 
but also the phonons. Including only the atoms our list is not complete! 
However, when you destroy the crystal you do not find the phonons! They 
disappear! On the other hand, if you want to reconstruct your crystal after 
you have broken it, the atoms you were left with are not enough: you must 
supplement the information which tells them to sit in the special lattice 
you want (cubic or else, etc.). You need, in short, to supplement the 
ordering information which was lost when the crystal was destroyed. Exactly 
such an ordering information is "dynamically" realized in the phonon 
particles. Thus, the phonon particle only exists (but really exists!) as 
long as the crystal exists, and vice versa. The function of being crystal 
is identified with the particle structure! As you see there is a lot in the 
quantum theory of matter and please notice: the description of crystal in 
terms of phonons has nothing to do with "interpretative problems". It is a 
well understood, experimentally well tested physical description.

Such a situation happens many times in physics; other familiar examples are 
ferromagnets, superconductors, etc.. It is a general feature occurring when 
the symmetry of the dynamics is not the symmetry of the states of the 
system (symmetry is spontaneously broken, technically speaking). Let me 
explain what this means. Consider the crystal as an example: the symmetry 
of the dynamics is the continuous space translational symmetry (the atoms 
may move around occupying any position in the available space). In the 
crystal state however such a symmetry is lost (broken) since the atoms must 
get ordered in the lattice sites; they cannot sit, e.g., in between two 
lattice corners: order is lack of symmetry!
A general theorem states that when a continuous symmetry is spontaneously 
broken, or equivalently, as we have just seen, an ordered pattern is 
generated, a massless particle is dynamically created; this particle 
(called the Nambu-Goldstone boson) is the phonon in the crystal case. 
Please, notice that this particle is massless, which means that it can span 
the whole system volume without inertia, which in turn guaranties that the 
ordering information is carried around without losses and that the ordered 
pattern is a stable one since the presence (or, as we say, the 
condensation) of the Goldstone particles of lowest momentum does not add 
energy to the state (it is enough to consider the lowest energy state, 
namely the ground state); in conclusion, the ordered ground state has the 
same energy of the symmetric (unordered) one (we call it normal ground 
state): they are degenerate states. This is why the crystal does exist as a 
stable phase of the matter. Actually, ground states, and therefore the 
phases the system may assume, are classified by their ordering degree (the 
order parameter) which depends on the condensate density of Goldstone 
quanta. We thus see that by tuning the condensate density (e.g. by changing 
the temperature) the system may be driven through the phases it can assume. 
Since the system phases are macroscopically characterized (the order 
parameter is in fact a macroscopic observable), we see that a bridge 
between the microscopic quantum scale and the macroscopic scale is 
established.

All the above is of course possible only if the mathematical formalism 
provides us with many degenerate but physically inequivalent ground states 
which we need to represent the system phases, which in fact have different 
physical properties: this is why we have to use QFT and not QM, as I said 
above. In QM all the possible ground states are physically equivalent (the 
Von Neuman Theorem); QFT is on the contrary much richer, it is equipped 
with infinitely many, physically inequivalent ground states and therefore 
we must use QFT to study systems with many phases.

Above I have been mentioning "theorems": however, I want to stress that 
these mathematical theorems perfectly fit and are fitted by real 
experiments and they represent the only available quantum theory (QFT 
indeed) on which the reliable working of any sort of technological gadget 
around us is based; in spite of the many epistemological and philosophical 
unsolved questions quantum theories may arise.

Now you see why I said that I need to start by considering actual material: 
this is not simply a list of constituents, it is not simply specific 
information from punctual observations, it is not simply a lot of real data 
and statistics, but it is also the dynamics. Otherwise, I would only be 
like one of those extremely patient and skillful swiss watch-makers who in 
the past centuries by mechanically assembling together a lot of wheels and 
levers and hooks were building beautiful puppets able to simulate many 
human movements. But... the phone book is not enough and we know that it 
CANNOT even be complete without the dynamics. There is no hope to build up 
a crystal without the long range correlations mediated by the phonons: if 
you try to fix up atom by atom in their lattice sites holding them by hooks 
you will never get the coherent orchestra of vibrating atoms playing the 
crystal function. This is what experiments tell us.

For every new or more refined movement more and more specialized units and 
wheels were needed in building the eighteenth century puppets. And 
certainly the brain, and living matter in general, do present a lot of very 
specialized units, which we absolutely need to search for. But our list of 
components will still possibly be incomplete if we do not make the effort 
of thinking of a dynamical scheme, too. There are properties of living 
matter, such as self-ordering, far from the equilibrium behaviour, 
non-dissipative energy transfer on protein molecular chains and at the same 
time dissipativity of biological systems, extremely high chemical 
efficiency and at the same time extremely high number of chemical species, 
and so on, that do point to the existence of a non-trivial dynamical 
background out of which the rich phenomenology of molecular biology 
emerges. Like with chemistry before the birth of QM, we are challenged to 
search for a unifying dynamical scheme, which may help us in understanding 
those (collective) properties not in the reach of the assembly by "hooks" 
of the units listed in our phone book.

The problem is not why to expect a quantum dynamical level in living matter 
(and in the brain). In its "inert (or dead!) phase" the matter counts among 
its components atoms, molecules and, as we have seen, other units 
dynamically generated (e.g. the phonon), all of them ruled by quantum laws. 
It would be a really crazy world the one where the same atoms, molecules 
and dynamically generated units would not be ruled by the same quantum laws 
in the "living phase" of matter.

Sometime people gets confused between "classical level" and "quantum 
level". We do speak about "classical limit" of quantum physics, but we 
NEVER mean that, e.g., the Planck constant "becomes" (or "goes" to) zero in 
the classical limit (even when, for sloppiness, we do say that; sorry!). 
The Planck constant has a well definite value which NEVER is zero! By 
"classical" we only mean that certain properties of the system are 
acceptably well described, from the observational point of view, "in the 
approximation" in which certain ratios between the Planck constant and some 
other quantity (of the same physical dimensions) are neglected. This does 
not mean that in such a case one "puts" the Planck constant equal to zero, 
because there are other behaviours, which the same system shows 
simultaneously to the "classical" ones, which only can be described by 
keeping the non-zero value of the Planck constant in its full glory. An 
example: our friend the crystal does certainly behaves as a classical 
object in many respects, out of any possible doubt. However, the 
phonon IS a quantum particle and therefore the macroscopic 
function of being a crystal IS a quantum feature of our system; not only, 
but it is indeed such a quantum behaviour, the one of being a crystal, that 
allows the "classical" behaviour of the components atoms as a "whole". 
Therefore, a diamond is a macroscopic quantum system classically behaving 
when one gives it as a gift to his/her fiance' (and let's hope they will 
not argue about the phonon, the Schroedinger cat, their love being 
classical or quantum and all that; it would be not at all romantic!).

In the same way, systemic features of living matter, such as ordered 
patterns, sequentially interlocked chemical reactions, non-dissipative 
energy transfer, nonlocal simultaneous response to external stimuli, etc., 
may result as macroscopic quantum features supporting the rich 
phenomenology of molecular biology: the idea, in the QFT approach to living 
matter, is to supplement with a basic dynamics the phenomenological random 
kinematics of biochemistry. 

So the problem is not "if" there exist a quantum dynamics in living matter 
(how it could not exist!), but which are its observable manifestations, if 
any, and in any case how the biochemistry as it is emerges from it. Of 
course, it is more and more urgent the need to know all what we can know 
about the components, their kinematics, their engineering; we need working 
models to solve immediate problems (floating boats were used well before 
knowing Archimede's law); we even need patient assembly of cells by hooks 
to form a tissue, but we cannot cry at sky if a cancer develops: from the 
hook strategy point of view only random kinematics and no dynamics is 
involved in tissue formation and as a consequence there is no reason why 
the same list of component cells should behave as a tissue instead of as 
a cancer. Sometime also the eighteenth century puppets were falling down 
in pieces. Therefore, it might be worthwhile to apply what we have 
learned about collective modes holding up atoms in the lattice sites (the 
crystal is a "tissue"!), spontaneous symmetry breakdown, coherence, boson 
condensation, etc., to study, together with biochemists and biologists, 
e.g., the "normal" (or symmetric) state of cancer and the ordered state 
of tissue, as we would say in QFT language.

The task is not at all simple. Living matter is not an inert crystal. And 
we should expect many surprises. For example in the quantum model of the 
brain by Umezawa and Ricciardi the problem of memory capacity seems to be 
solved by seriously considering the dissipative character of the brain 
system. That dissipation enters into play can be naively understood by 
observing that information recording breaks the symmetry under time 
reversal, i.e. it introduces the arrow of time: "NOW you know it...!" is 
the warning to mean that "after" having received some information, one 
cannot anymore behave as "before" receiving it. Thus memorizing breaks time 
reversal symmetry. The brain dynamics is therefore intrinsically 
irreversible. In more familiar words, the brain, as other biological 
systems, has a history. In this respect the brain is a clock. Well, to 
treat dissipative brain dynamics in QFT one has to introduce the 
time-reversed image of the system degrees of freedom. One finds thus 
himself dealing with a system made by the brain and by its "mirror in 
time" image, as a result of the internal consistency of the mathematical 
scheme (if you want to know more about that look at my paper in 
Int.Journal of Mod. Phys. B9 (1995) 973).
Problem: are consciousness mechanisms macroscopic manifestations of the 
mirror brain dynamics? Does the conscious experience of the flow of time 
emerges from the brain dissipative dynamics? The mirror modes are related 
to brain-environment coupling and at the same time to brain 
self-interaction. Does this lead to the conscious sense of "self"?

I realize this is a long letter and I will not talk any longer about 
brain and living matter, consciousness and QFT. I stop here, otherwise the 
Editors of the book on Tucson II will complain for the exceeding number of 
words and I risk to be left out as it was for Tucson I book. I hope we can 
resume our discussion in a future occasion in order to be able to join our 
efforts in the study of the brain.

Arrivederci a presto,   Giuseppe

P.S. I thank you for allowing me to publish this letter. G.

\end{document}